\documentclass[superscriptaddress,showpacs,showkeys,preprintnumbers,amsmath,amssymb,aps,twocolumn,nofootinbib,prd]{revtex4-1}
\usepackage{graphicx}
\usepackage{epsfig}

\begin{document}
\title{\textbf{Stability analysis of de Sitter solutions in models with the Gauss--Bonnet term}}

\author{Ekaterina~O.~Pozdeeva}
\email{pozdeeva@www-hep.sinp.msu.ru}
\affiliation{Skobeltsyn Institute of Nuclear Physics, Lomonosov Moscow State University,\\
 Leninskie Gory 1, Moscow  119991,  Russia}

\author{Mohammad~Sami}
\email{msami@jmi.ac.in}  \affiliation{Centre
for Theoretical Physics, Jamia Millia Islamia, New Delhi 110025,
India}
\affiliation{Maulana Azad National Urdu
University, Gachibowli, Hyderabad 500032, India}
\affiliation{Institute for Advanced Physics and Mathematics, Zhejiang University of Technology,\\ Hangzhou 310032, China}

\author{Alexey~V.~Toporensky}
\email{atopor@rambler.ru}
\affiliation{Sternberg Astronomical Institute, Lomonosov Moscow State University, Universitetsky pr.~13, Moscow 119991, Russia}
\affiliation{Kazan Federal University, Kremlevskaya Street~18, Kazan 420008, Russia}

\author{Sergey~Yu.~Vernov}\email{svernov@theory.sinp.msu.ru}
\affiliation{Skobeltsyn Institute of Nuclear Physics, Lomonosov Moscow State University,
 Leninskie Gory 1, Moscow  119991,  Russia}

\begin{abstract}
     We investigate the scalar field dynamics of models with nonminimally coupled scalar fields in the presence of the Gauss--Bonnet term and derive the structure of the effective potential and conditions for stable de Sitter solutions in general.
  Specializing to specific couplings, we explore the possibility of realizing the stable de Sitter configurations which may have implications for both the early Universe and late time evolution.
  \end{abstract}
\maketitle

\section{Introduction}
Among curvature corrections to the Einstein--Hilbert action, the Gauss--Bonnet term is of a special relevance. The Gauss--Bonnet  term is a topological invariant quantity in four  dimensions but may have dynamical relevance if is coupled to an evolving scalar field. The Gauss--Bonnet  term coupled to a scalar field arises naturally in the string theory~\cite{String} and uses in string inspired cosmological models~\cite{Antoniadis:1993jc,Torii:1996yi,Kawai1998,Hwang:2005hb,Tsujikawa:2006ph,ss}, because has interesting cosmological implications both for early Universe~\cite{vandeBruck:2015gjd,vandeBruck:2016xvt,Wu:2017joj,Nozari:2017rta,Guo:2010jr,Oikonomou:2017ppp,Odintsov:2018zhw,Fomin:2019yls} and for the late time  dynamics~\cite{Nojiri:2005vv,Sami:2005zc,Cognola:2006eg,Cognola:2006sp,Elizalde:2010jx,Benetti:2018zhv}\footnote{Nonlocal cosmological models with the Gauss--Bonnet  term are actively developed as well~\cite{Capozziello:2008gu,Cognola:2009jx,Koshelev:2013lfm,Buoninfante:2018xiw,Elizalde:2018qbm,Tian:2019bla}.}.

In the  case of the scalar field having to account for late time cosmic evolution, it is more than desirable to ensure that it does not disturb the thermal history and that the late time dynamics is free from initial conditions of the field.
In this case, it is mandatory that dynamics exhibits scaling behavior where the scalar field, in a sense, spends most of its time. In that case, the thermal history remains intact. However, a scaling solution, which is an attractor of dynamics,
mimics background matter, and one, therefore, needs to devise an exit mechanism  from this regime to acceleration. In the case of the Einstein--Hilbert action, the steep exponential potential with the exponential form of the coupling function gives rise to an exit from the scaling regime to acceleration~\cite{Tsujikawa:2006ph,ss}. Indeed, the Gauss--Bonnet coupling induces a minimum in the runaway potential where the field can settle such that the de Sitter is a late time attractor of the dynamics~\cite{Tsujikawa:2006ph,ss}. Let us note that there exist other mechanisms capable of executing the said transition. For instance, coupling with massive neutrino matter to a scalar field proportional to a trace of the energy-momentum tensor of neutrino matter can facilitate an exit from scaling regime. In this case, coupling dynamically builds up only at late stages, resulting in the minimum of the effective potential~\cite{sss}. In general, the addition of the Gauss--Bonnet term modifies the field potential and may induce interesting features in~it~\cite{Tsujikawa:2006ph,ss}.

In order to find out the de Sitter solutions in a model with a minimally coupled scalar field with a potential $V$, it is enough to find zeros of the first derivative of $V$. The sign of the second derivative of the potential $V$ at a de Sitter point determines the stability of the solution.  Similar analysis of de Sitter solutions is possible in the case of nonminimal coupling if one introduces an effective potential such that zeroes of its first derivative correspond to de Sitter points whereas the sign of its second derivative determines their stability~\cite{Skugoreva:2014gka,Pozdeeva:2016cja}.

In this paper, we  analyze  models with  the Gauss--Bonnet term in  case of a nonminimally coupled scalar field. We look for the general structure of the corresponding effective potential and investigate the specific cases. Using the framework of the effective potential, our  goal is to find out de Sitter solutions in these models and compare them with their counterparts in the corresponding models without the Gauss--Bonnet term.

The paper is organized as follows.
In Section~II, we remind evolution equations for models with a nonminimally coupled scalar field and the Gauss--Bonnet term. In Section~III, we propose the effective potential as a useful tool for the search and stability analysis of de Sitter solutions. In Section~IV,  we present a few specific examples of  models with de Sitter solutions. Section~V is devoted to our conclusions.

\section{Evolution equations for models with the Gauss--Bonnet term}
In this paper, we  investigate the  model with the Gauss--Bonnet term in a general background described by the following action:
\begin{equation}
\label{actionl}
S=\!\int\! d^4 x\sqrt{-g}\left[UR-\frac{1}{2}g^{\mu\nu}\partial_{\mu}\phi\partial_{\nu}\phi- V-F\mathcal{G}\right],
\end{equation}
where the functions $U(\phi)$, $V(\phi)$, and $F(\phi)$ are differentiable ones, and $\mathcal{G}$ is the Gauss--Bonnet term,
\begin{equation*}
\mathcal{G}=R^2-4R_{\mu\nu}R^{\mu\nu}+R_{\mu\nu\alpha\beta}R^{\mu\nu\alpha\beta}.
\end{equation*}

In what follows, we specialize to the spatially flat Friedmann--Lema\^{i}tre--Robert\-son--Walker universe,
\begin{equation}
\label{metric}
ds^2={}-dt^2+a^2(t)\left(dx_1^2+dx_2^2+dx_3^2
\right).
\end{equation}

Variations of action (\ref{actionl}) with respect to $g_{\mu\nu}$ and $\phi$ lead to the following evolution equations~\cite{vandeBruck:2015gjd}:
\begin{equation}\label{Equ00}
    6H^2U+6HU'\dot{\phi}=\frac{1}{2}{\dot{\phi}}^2+V+24H^3F'\dot{\phi},
\end{equation}
\begin{equation}\label{EquH}
    4\left(U-4H\dot{F}\right)\dot{H}={}-\dot{\phi}^2-2\ddot{U}+2H\dot{U}+8H^2\left(\ddot{F}-H\dot{F}\right),
   \end{equation}
\begin{equation}
\label{equphi}
\ddot\phi+3H\dot\phi-6\left(\dot H+2H^2\right)U'+V'+24H^2F'\left(\dot H +H^2\right)=0,
\end{equation}
where  $H=\dot{a}/a$ is the Hubble parameter, dots and primes denote the derivatives with respect to the cosmic time and  $\phi$, respectively.

It would be convenient to cast Eqs.~(\ref{EquH}) and (\ref{equphi}) as a dynamical system,
\begin{equation}
\label{DynSYS}
\begin{split}
\dot\phi=&\psi,\\
\dot\psi=&\frac{1}{2\left(B-4F'H\psi\right)}\left\{2H\left[3B+4F'V'-6{U'}^2-6U\right]\psi
\right.\\
-&\left. 2\frac{V^2}{U}X+ \left[12\left[\left(2U''+3\right)F'+2U'F''\right]H^2
\right.\right.\\
-&\left.\left.
96F'F''H^4-3\left(2U''+1\right)U'\right]\psi^2 \right\},\\
\dot H=&\frac{1}{4\left(B-4F'H\psi\right)}\left\{8\left(U'-4F'H^2\right)H\psi\right.\\
-&\left. 2\frac{V^2}{U^2}\left(4F^\prime H^2-U'\right) X+ \left(8F''H^2-2{U''}-1\right)\psi^2\right\},
\end{split}
\end{equation}
where
\begin{equation}
\label{B}
B=3\left(4H^2F^\prime -U^\prime \right)^2+U,
\end{equation}
\begin{equation}
\label{X}
X=\frac{U^2}{V^2}\left[24F^\prime  H^4-12 U^\prime  H^2+V^\prime \right].
\end{equation}
Dynamical system (\ref{DynSYS}) is suitable for the search for de Sitter solutions, which we shall undertake in the following section.

\section{The effective potential and de Sitter solutions}
\subsection{The search for de Sitter solutions}
Our first goal is to find de Sitter solutions in the model described by action (\ref{actionl}) and to compare them with de Sitter solutions in the corresponding model without the Gauss--Bonnet term. In principle, solutions with a constant Hubble parameter $H$ and the field $\phi$ depending on time can exist, but these solutions correspond to a varying effective gravitational constant, which is proportional to $1/U(\phi)$, and have properties different from the standard de Sitter solutions. For this reason, we restrict ourselves to de Sitter solutions with a constant~$\phi$.
Substituting $\phi=\phi_{dS}$ and $H=H_{dS}$ into Eqs.~(\ref{Equ00}) and (\ref{equphi}), we have
\begin{equation}
\label{Equ00dS}
    6H_{dS}^2U_{dS}=V_{dS},
\end{equation}
\begin{equation}
\label{equphidS}
12H_{dS}^2U'_{dS}=V'_{dS}+24H_{dS}^4F'_{dS},
\end{equation}
where $V_{dS}=V(\phi_{dS})$, $U_{dS}=U(\phi_{dS})$, and $F_{dS}=F(\phi_{dS})$.
We see that the value of the Hubble parameter at the de Sitter point is the same as in the corresponding model without  the Gauss--Bonnet term,
\begin{equation}
\label{HdS}
H_{dS}^2=\frac{V_{dS}}{6U_{dS}}.
\end{equation}
The value of $F'(\phi_{dS})$ is fixed by Eq.~(\ref{equphidS}):
\begin{equation}
\label{equdS}
F'_{dS} = \frac{3U_{dS}(2U'_{dS}V_{dS}-V'_{dS}U_{dS})}{2V_{dS}^2}.
\end{equation}
Therefore, we come to the conclusion that for arbitrary functions $U$ and $V$ with  $VU>0$, we can choose  $F(\phi)$ such that the corresponding point becomes a de Sitter solution, with the Hubble parameter defined by Eq.~(\ref{HdS}).

Let us note that at a de Sitter point, the system of equations~(\ref{DynSYS}) has the following form:
\begin{eqnarray}
\dot{\phi}_{dS}&=&0,\\
\dot{\psi}_{dS}&=&{}-\frac{V_{dS}^2}{B(\phi_{dS})U_{dS}}X(\phi_{dS})=0,\\
\dot{H}_{dS}&=&{}-\frac{\left(4F_{dS}^\prime H_{dS}^2-U^\prime_{dS}\right)V_{dS}^2 }{2B(\phi_{dS})U_{dS}^2}X(\phi_{dS})=0.
\end{eqnarray}

It would be convenient if all the necessary information on the existence and stability of de Sitter solutions is obtained  from a single combination of functions $U$, $V$, and $F$ dubbed effective potential $V_{eff}$. The  de Sitter solutions would correspond to zeros of the first derivative of $V_{eff}$, and the
stability of the solutions would correspond to   its second derivative being positive.

 Let us emphasize that we require only these two conditions for the definition of the effective potential: this is a function $V_{eff}(\phi)$ such that $V_{eff}'=0$ at de Sitter
point $\phi_{dS}$, and this de Sitter solution is stable if the second derivative of the effective potential
with respect to $\phi$ is positive for $\phi=\phi_{dS}$.  Such a definition is a direct generalization of a scalar field potential $V(\phi)$  in the theory with a minimally coupled scalar field, since, in this theory, the de Sitter solution requires $V'=0$ and is stable in the
point where the scalar field potential reaches its local minimum. We do not require that any
other properties of the physical system considered should be expressed  through properties
of the effective potential.

In the present paper, we show that while the general situation is more complicated, we achieve the above-mentioned goal
in the case of~$U>0$.

De Sitter solutions in the model with nonminimal coupling and without the Gauss--Bonnet term have been considered in~\cite{Skugoreva:2014gka}, where the effective potential has been introduced. In the model with the Gauss--Bonnet term, we  can also introduce such $V_{eff}(\phi)$
that
\begin{equation}
\label{DVeff}
V'_{eff}(\phi_{dS})=0.
\end{equation}

Indeed, let
\begin{equation}
\label{V_eff}
    V_{eff}={}-\frac{U^2}{V}+\frac{2}{3}F.
\end{equation}

Using Eq.~(\ref{HdS}), we get
\begin{equation}\label{XdS}
    X(\phi_{dS})=\frac{2}{3}F^\prime_{dS}-2 \frac{U^\prime_{dS}U_{dS}}{V_{dS}}+\frac{V^\prime_{dS} U_{dS}^2}{V^2_{dS}}=V'_{eff}(\phi_{dS}),
\end{equation}
therefore, Eq.~(\ref{DVeff}) is satisfied, and  de Sitter solutions correspond to extremum points of the effective potential~$V_{eff}$.

It should be noted that the first term in Eq.~(\ref{V_eff}) can be expressed through the effective potential without
the Gauss--Bonnet term $\tilde V_{eff}$ introduced in~\cite{Skugoreva:2014gka} as $-\tilde V_{eff}^{-1}$.
A disadvantage of the form (\ref{V_eff}) for the effective potential is that it diverges at $V=0$.
However,  in such a case we have a Minkowski solution instead of de Sitter one, so this feature
does not spoil the analysis of de Sitter solutions.

Note that the effective potential $V_{eff}$ (\ref{V_eff})
is not a unique function suitable to define the values of $\phi_{dS}$ and the stability of de Sitter solutions. For example, one can use
\begin{equation}
\label{V2eff}
    \tilde{V}_{eff}={}-\frac{1}{V_{eff}}=\frac{3V}{3U^2-2FV}\,,
\end{equation}
as an effective potential.
It is easy to see that $\tilde{V}_{eff}'(\phi_{dS})=0$, and the sign of $\tilde{V}_{eff}''(\phi_{dS})$ coincides with the sign of $V_{eff}''(\phi_{dS})$.
In the case of the absence of the Gauss--Bonnet term, $F=0$, $\tilde{V}_{eff}$ is the effective potential, introduced in Ref.~\cite{Skugoreva:2014gka}. Moreover, the function $\tilde{V}_{eff}$ is proportional to $V(\phi)$ in the case of $F=0$ and a constant $U$.

\subsection{The stability analysis}

To investigate the Lyapunov stability of a de Sitter solution we use the following expansions:
\begin{eqnarray*}
H(t)&=&H_{dS}+\delta H_1(t),\\ \phi(t)&=&\phi_{dS}+\delta \phi_1(t),\\ \psi(t)&=&\delta \psi_1(t),
\end{eqnarray*}
where $\delta$ is a small parameter. From (\ref{DynSYS}), in the first order in $\delta$ we get the following linear system
\begin{eqnarray}
  \dot{\phi}_1&=&A_{11}\phi_1+A_{12}\psi_1+A_{13}H_1 \,,\label{dphi1}\\
  \dot{\psi}_1&=&A_{21}\phi_1+A_{22}\psi_1+A_{23}H_1\,,\label{dpsi1}\\
   \dot{H}_1&=&A_{31}\phi_1+A_{32}\psi_1+A_{33}H_1\,,\label{dH1}
\end{eqnarray}
where
\begin{widetext}
 \begin{equation*}
 A=
 \begin{array}{||ccc||}
   0, & 1, & 0 \\
   &&\\
   {}-\frac{V^2}{UB}X'_{,\phi}\,,\qquad & H_{dS}\left(1-4\frac{U}{B}\right),\qquad & {}-\frac{V^2}{UB} X'_{,H} \\
   &&\\
   {}\frac{V X'_{,\phi}}{2BU^2}\left(V'U-U'V\right),\qquad & \frac{2H_{dS}}{BV}\left(V'U-U'V\right),\qquad & \frac{V X'_{,H}}{2BU^2}\left(V'U-U'V\right)
 \end{array}
 \end{equation*}
and all functions are taken at $\phi=\phi_{dS}$.
\end{widetext}

Since $\det(A)=0$, functions $H_1(t)$, $\phi_1(t)$ and $\psi_1(t)$ are not independent. From Eq.~(\ref{Equ00}), we obtain
\begin{equation}\label{H1}
    H_1=\frac{V'_{dS}U_{dS}-U'_{dS}V_{dS}}{2U_{dS}V_{dS}}\left(H_{dS}\phi_1-\psi_1\right).
\end{equation}

Substituting \eqref{H1} into \eqref{dphi1} and \eqref{dpsi1}, we get:
\begin{eqnarray}
   \dot{\phi}_1&=&\tilde{A}_{11}\phi_1+\tilde{A}_{12}\psi_1\,,\label{Dphi1}\\
  \dot{\psi}_1&=&\tilde{A}_{21}\phi_1+\tilde{A}_{22}\psi_1\,,\label{Dpsi1}
\end{eqnarray}
where
\begin{widetext}
\begin{equation*}
\tilde{A}=\begin{array}{||cc||}0,&1\\{}
-{\frac {{V}
^{2}{X^\prime_{,\phi}}}{  U B}}-{\frac { V{
X^\prime_{,H}}\, \left( {V^\prime}\, U -{U^\prime}\, V \right) H_{dS}}{2{U}^{2}B}},\qquad &
 H_{dS}\left(1-4\frac{U}{B}\right)+{\frac {V {X^\prime_{,H}}\, \left( V^\prime\, U -U^\prime\, V  \right) }{2{U}^{2}B}}
\end {array}\,.
\end{equation*}
\end{widetext}

Solutions of Eqs.~(\ref{Dphi1}) and (\ref{Dpsi1}) have the following form:
\begin{equation}\label{solphi1}
  {\phi}_1=c_{11}\mathrm{e}^{-\lambda_{-}t} +c_{21}\mathrm{e}^{-\lambda_{+}t},
\end{equation}
\begin{equation}
\label{solpsi1}
  {\psi}_1=c_{21}\mathrm{e}^{-\lambda_{-}t} +c_{22}\mathrm{e}^{-\lambda_{+}t},
\end{equation}
 where $c_{ij}$ are some constants whose values are not important for the stability analysis. The numbers $\lambda_\pm$ are roots of
the determinant of the characteristic matrix. Solving the equation
\begin{equation}
\det(\tilde{A}-\lambda\cdot I)=0,
\end{equation}
we get:
\begin{equation}\label{lambda12}
   \lambda_\pm=\frac{Z\pm\sqrt{Z^2+Y}}{4U^2B}\,,
\end{equation}
where
\begin{equation*}
Z={}-\frac{3U^2}{V^2}\sqrt{\frac{6V}{U}}\left[\frac{7}{9}UV^2+(V'U-U'V)^2\right],
\end{equation*}
and
\begin{equation*}
\begin{split}
Y&=8VB\!\left[{X^\prime_{,H}}\,H_{dS}{U}^{
2}V{U^\prime}-{X^\prime_{,H}}\,H_{dS}{U}^{3}  {V^\prime}-2{U}^{3}{V}{X^\prime_{,\phi}}\right]\\
&={}-16U^3V^2BV''_{eff}\,.
\end{split}
\end{equation*}

To get this result, we assume that $H_{dS}>0$, because only such solutions correspond to the observable Universe, and put $H_{dS}=\sqrt{\frac{V}{6U}}$.

A de Sitter solution is stable if real parts of both $\lambda_-$ and $\lambda_+$ are negative.  In the case of a positive $U$ and, therefore, a positive $V$, we see that $Z<0$ and $B>0$. This means that the condition
$Z/B<0$  is satisfied at any de Sitter point, and, therefore, $\Re e (\lambda_-)<0$.

At $U>0$, the condition $\Re e (\lambda_+)<0$ is equivalent to $Y<0$ and, hence, $V''_{eff}(\phi_{dS})>0$.
Thus, we finally reach a conclusion that for any $U(\phi_{dS})>0$, a de Sitter solution is stable if $V''_{eff}(\phi_{dS})>0$ and unstable if $V''_{eff}(\phi_{dS})<0$.

In the next section, we consider several examples of models  and explore the existence and stability of  de Sitter solutions.

\section{Examples}

\subsection{Models with exponential potential}

The string theory inspired cosmological model with
\begin{equation}
\label{modelexp}
    U=U_0,\quad V=c\mathrm{e}^{-\tilde{\lambda}\phi},\quad F=\frac{\alpha}{\mu}\mathrm{e}^{\mu\phi},
\end{equation}
where $U_0$, $\alpha$, $c$, $\tilde{\lambda}$, and $\mu$ are positive constants, has been considered in~\cite{Tsujikawa:2006ph}.
In this model, the effective potential is
\begin{equation}
\label{Veffexp}
    V_{eff}={}-\frac{U_0^2}{c}\mathrm{e}^{\tilde{\lambda}\phi}+\frac{2\alpha}{3\mu}\mathrm{e}^{\mu\phi}.
\end{equation}
The condition $V_{eff}'(\phi_{dS})=0$ gives
\begin{equation}
\label{expdS}
    \phi_{dS}=\frac{1}{\tilde{\lambda}-\mu}\ln\left(\frac{2ac}{3U_0^2\tilde{\lambda}}\right).
\end{equation}
There exists a de Sitter solution for all $\mu\neq\tilde{\lambda}$. It is easy to see that $V_{eff}^{\prime\prime}=0$ at
\begin{equation}
    \phi_2 = \frac{1}{\tilde{\lambda}-\mu}\ln\left(\frac{2ac\mu}{3U_0^2\tilde{\lambda}^2}\right)
    =\phi_{dS}-\frac{\ln(\tilde{\lambda})-\ln(\mu)}{\tilde{\lambda}-\mu}\,,
\end{equation}
and $\phi_{dS}>\phi_2$ for any $\tilde{\lambda}\neq \mu$.

If $\mu>\tilde{\lambda}$, then $V_{eff}^{\prime\prime}$ is positive at large $\phi$, so the second derivative is positive at the de Sitter point and this point is stable. In the opposite case,  $\mu<\tilde{\lambda}$, $V_{eff}^{\prime\prime}<0$ at large $\phi$ and the de Sitter solution is unstable. This result coincides with the result obtained in~\cite{Tsujikawa:2006ph} by another method.

We generalize this result assuming that the constants can be negative. For
\begin{equation}\label{Ex}
    {V_{eff}}=c_1\mathrm{e}^{-N_1\phi}+c_2\mathrm{e}^{-N_2\phi},
\end{equation}
the de Sitter point $\phi_{dS}={\frac {1}{N_1-N_2}\ln  \left({} -{\frac {c_1\,{N_1}}{c_2\,{N_2}}} \right) }$ exists only if
${\frac {c_1\,{N_1}}{c_2\,{N_2}}}<0$ and $N_2\neq N_1$. If we assume functions  $U$ and $V$ in the form given by \eqref{modelexp}, then $c_1<0$.
At the same time, one and the same the effective potential corresponds to a different choice of functions $F$, $V$, and $U$.
If two of these functions are given, then we can get the third function using the given form of the effective potential.
It is a way of constructing models with de Sitter solutions. For example, the model with a nonminimally coupled scalar field defined by functions
\begin{equation*}
U={U_0}\left( \xi{\phi}^{2}+1 \right){\mathrm{e}^{\eta_1\,\phi}},\quad \mbox{and}\quad V=V_0\phi^4\mathrm{e}^{\eta_2\phi},
\end{equation*}
 has the effective potential
given by (\ref{Ex}) if
\begin{equation*}
F=\frac{3}{2}\left[{\frac {4U_0^{2}{\mathrm{e}^{2\,\eta_1\phi-\eta_2
\phi}}}{{V_0}}}\left(\xi+\frac{1}{\phi^2}\right)^2+c_1{\mathrm{e}^{-{N_1}\phi}}+c_2{\mathrm{e}^{-{N_2}\phi}}\right]\!.
\end{equation*}

In this model, $c_i$ and $N_i$ are arbitrary constants. The analysis of the second derivative of $V_{eff}$ gives the following stability conditions:
\begin{itemize}
  \item if $c_1>0$ and $c_2>0$, then the de Sitter solution  is stable;
  \item if $c_1<0$ and $c_2<0$, then the de Sitter solution  is unstable;
   \item if $c_1>0$ and $c_2<0$, then
   the de Sitter solution is stable at  $|N_1|>|N_2|$ and unstable at $|N_1|<|N_2|$;
    \item if $c_1<0$ and $c_2>0$, then
   the de Sitter solution is stable at $|N_1|<|N_2|$ and unstable at $|N_1|>|N_2|$.
\end{itemize}

The effective potential can be used not only to simplify the analysis of the stability of de Sitter solutions in a given model, but also to construct a new model with de Sitter solutions.

\subsection{Models with $V=CU^2$}

Let us consider the case  $V=CU^2$, where $C$ is a positive constant.
In this case, a model without the Gauss--Bonnet  term transforms to a model with a constant potential in the Einstein frame.
 If the Gauss--Bonnet  term is presented, then the function $F(\phi)$ plays a role  of the effective potential, fully determining the position and stability of the de Sitter solutions, because
 \begin{equation}\label{Veff1}
    V_{eff}={}-\frac{1}{C}+\frac{2}{3}F.
\end{equation}
So, values of $\phi_{dS}$ satisfy the condition $F'(\phi_{dS})=0$. From Eq.~(\ref{lambda12}), it follows:
\begin{equation}
\label{lambdasol}
    {\lambda}_{\pm}={}-\frac{\sqrt{6CU}}{4}\pm\frac{\sqrt{6CU\left[9\left(3{U'}^2+U\right)-16CU^2F''\right]}}{12\sqrt{3{U'}^2+U}}.
\end{equation}
For $U(\phi_{dS})>0$, a de Sitter solution is unstable at $F''<0$ and stable at $F''>0$.

Note that the only difference between minimal and nonminimal coupling cases is that values of the Hubble parameter at  de Sitter points
\begin{equation*}
H_{dS}^2=\frac{C}{6}U(\phi_{dS}),
\end{equation*}
can be different if $U$ is not a constant.

 In what follows, we shall demonstrate the working of a de Sitter search algorithm through concrete examples.

For
$F={A_4}{\phi}^{4}+{A_2}{\phi}^{2}$,
de Sitter points defined by the condition $F'=0$ are
\begin{equation}
{\phi_{dS}}_{\pm}=\pm\sqrt{-\frac{A_2}{2A_4}},\quad {\phi_{dS}}_{0}=0.
\end{equation}

It is evident that ${\phi_{dS}}_{\pm}$ are real only if constants $A_2$ and $A_4$ have different signs.
The values of the second derivative of $F$ at the de Sitter points are
\begin{equation*}
    \left.F^{\prime\prime}\right|_{{\phi_{dS}}_{\pm}}={}-4A_2,\quad \left.F^{\prime\prime}\right|_{{\phi_{dS}}_{0}}=2A_2.
 \end{equation*}

\begin{figure}[h!tbp]
\includegraphics[width=0.72\linewidth]{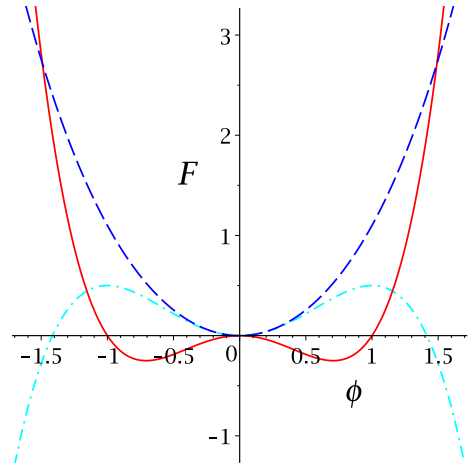} \
\caption{The function $F(\phi)={A_4}{\phi}^{4}+{A_2}{\phi}^{2}$ at different values of parameters: $A_2=1$ and $A_4=0.1$ (blue dash curve), $A_2=-1$ and $A_4=1$ (red solid curve), $A_2=1$ and $A_4=-0.5$ (cyan dash-dot curve).
\label{FigVU2}}
\end{figure}

 Thus, the de Sitter solution in points ${\phi_{dS}}_{\pm}$ is unstable for any $A_2>0$ and $A_4<0$ and is stable for any $A_2<0$ and $A_4>0$.
 At the point ${{\phi_{dS}}_{0}}$, the de Sitter solution is stable for any $A_2>0$ and unstable at $A_2<0$.
 At $A_2=0$, the only de Sitter point is $\phi_{dS}=0$ and we get  $ {\lambda}_{+}=0$ and  ${\lambda}_{-}={}-\sqrt{6CU}/2$.
 Figure~\ref{FigVU2} illustrates these different possibilities.

\begin{figure}[h!tbp]
\includegraphics[width=0.427\linewidth]{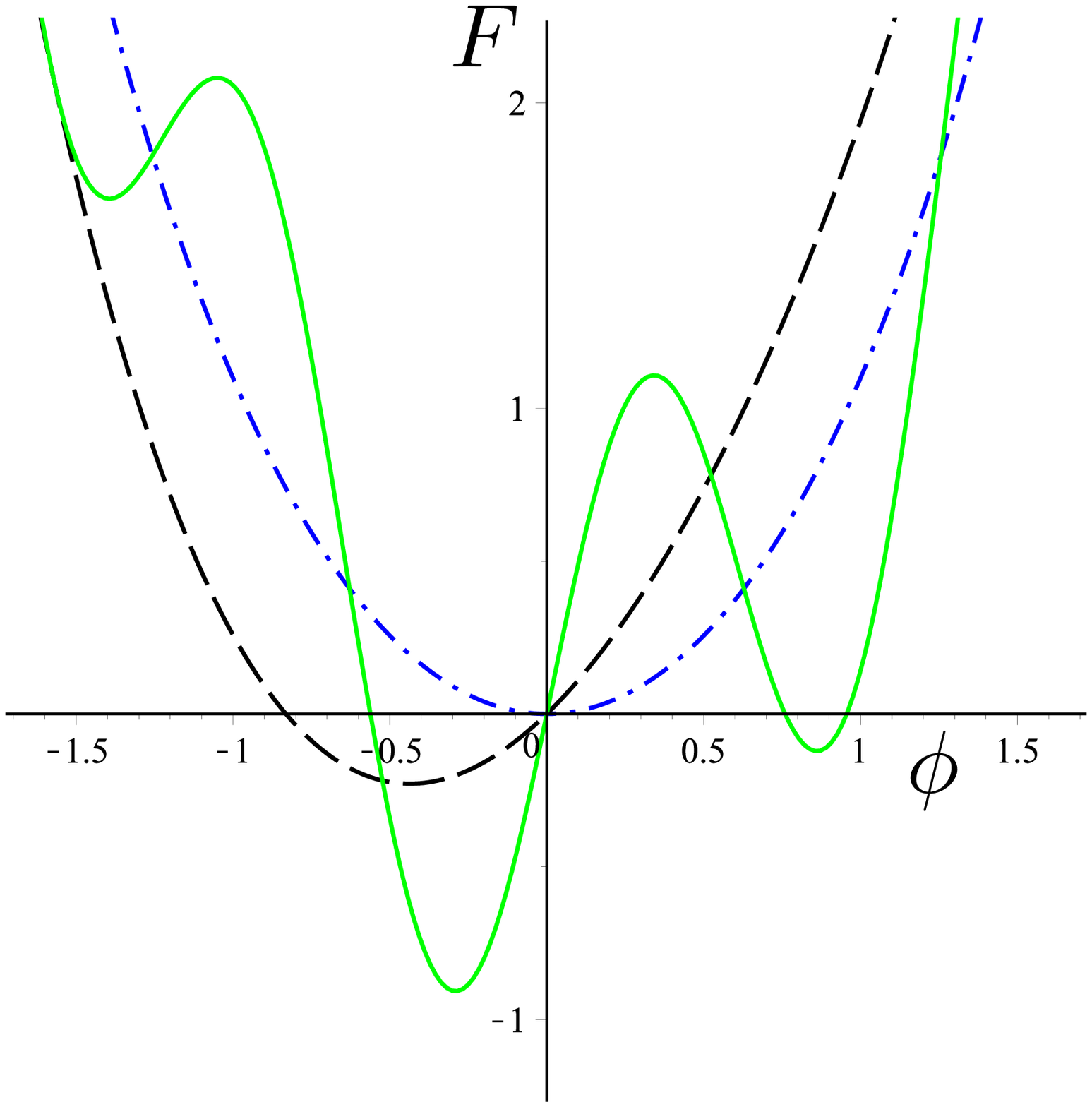} \
\includegraphics[width=0.427\linewidth]{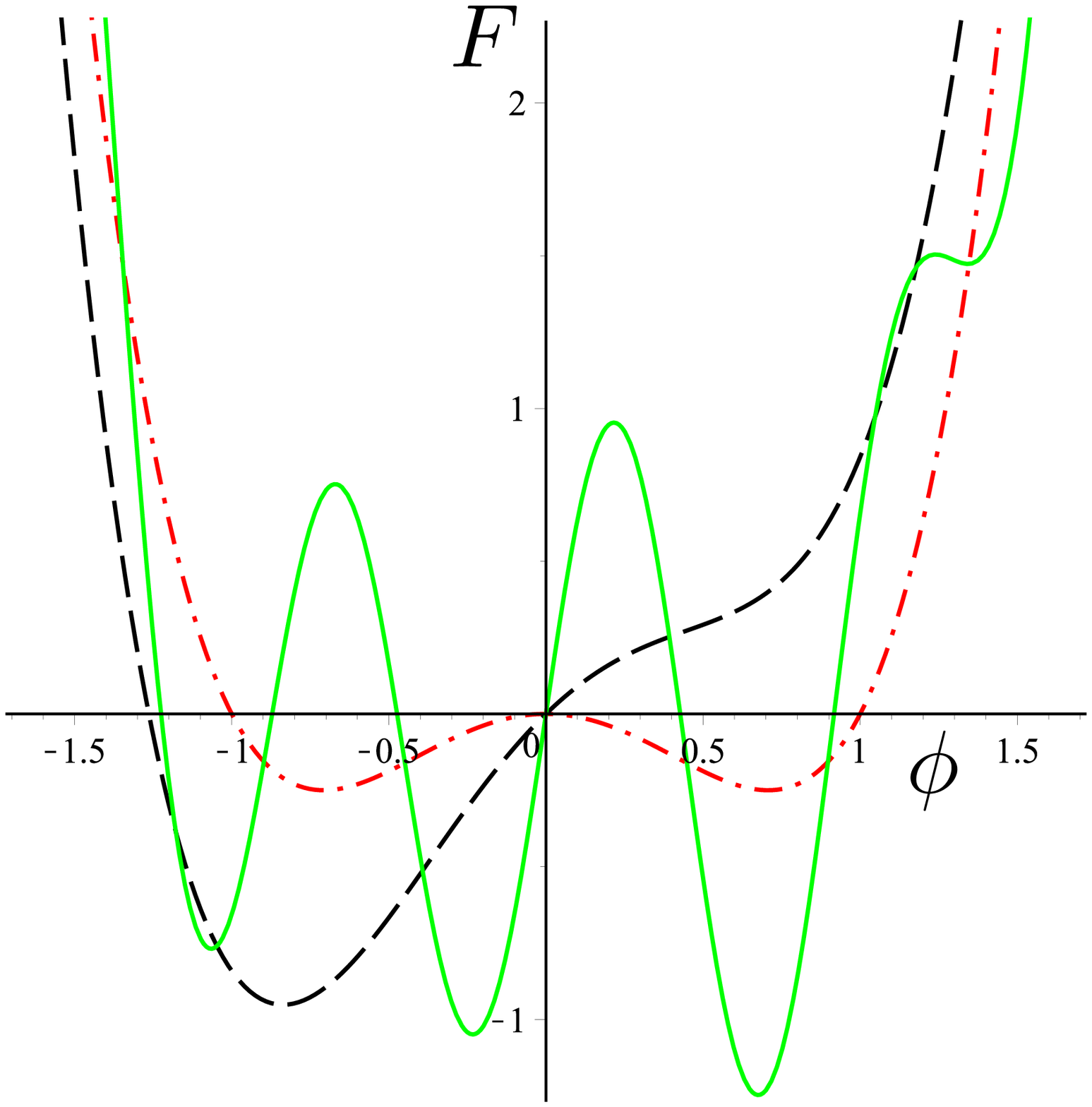}
\caption{The function $\tilde{F}(\phi)$ at different values of parameters. In the left picture, $A_2=1$, $A_4=0.1$, and $C=0$ (blue dash-dot curve), $C=1$ and $\omega=1$ (black dash curve), $C=1$ and $\omega=5$ (green solid curve).
In the right picture, $A_2=-1$, $A_4=1$, and $C=0$ (red dash-dot curve), $C=1$ and $\omega=1$ (black dash curve), $C=1$ and $\omega=7$ (green solid curve).
\label{FigVU2s}}
\end{figure}

A more complicated example is
\begin{equation}
\label{model3}
\tilde{F}={A_4}{\phi}^{4}+{A_2}{\phi}^{2}+C\sin(\omega\phi),
\end{equation}
where $C$ and $\omega$ are constants. Pictures in Fig.~\ref{FigVU2s} demonstrate that independence of values of the parameters  $C$ and $\omega$ the number of de Sitter solutions changes. In the left picture, the blue and black lines correspond to models with one stable de Sitter solution, whereas the green solid line corresponds to a model with 3 stable and 2 unstable de Sitter solutions. In the right picture, the red dash-dot curve corresponds to a model with 2 stable and 1 unstable de Sitter solutions, the black dash line corresponds to models with one stable de Sitter solution, and the green solid line corresponds to a model with 4 stable and 3 unstable de Sitter solutions. Therefore, using  a graphical representation of the function $\tilde{F}$, one can get the structure and stability properties of de Sitter solutions.

 Thus, for an arbitrary positive $U(\phi)$, we obtain de Sitter solutions given by an arbitrary function $V_{eff}$, if we choose  $V=CU^2$ and $F=V_{eff}$.

\subsection{Models with a massive scalar field}
In the previous subsection, the de Sitter solutions appear due to properties of the coupling function
$F$ solely. It is more interesting to get de Sitter solutions via an interplay between the scalar
field potential $V$ and the coupling function $F$ so these two functions, being simple monomials, give rise to a de Sitter solution. Using the conception of the effective potential, it is easy to create such models.
For example, let the potential be of the simplest massive form
\begin{equation}
V=m^2 \phi^2,
\end{equation}
with the coupling function
\begin{equation}
U=\xi \phi^2
\end{equation}
for the coupling function with curvature. In this situation, the effective potential is
\begin{equation}
V_{eff}={}-\frac{\xi^2}{m^2} \phi^2 + \frac{2}{3} F.
\end{equation}

\begin{figure}[h!tbp]
\includegraphics[width=0.72\linewidth]{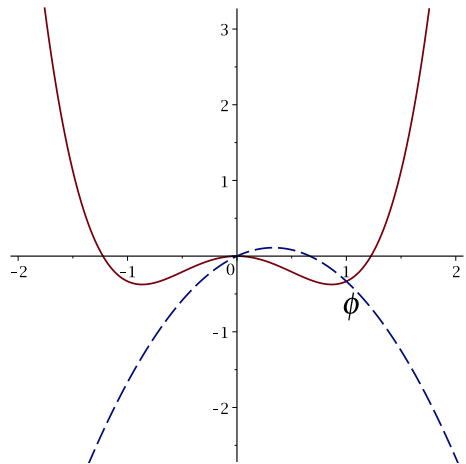}
\caption{The effective potential $V_{eff}(\phi)$ for $U=\phi^2$, $V=\phi^2$, $F=\phi^\alpha$
  is presented for $\alpha=4$ (red solid curve) and $\alpha=1$ (blue dash curve).
\label{Figpol2}}
\end{figure}

Without the Gauss--Bonnet contribution the effective potential is a monotonic function, so there are no de Sitter solutions\footnote{In the point $\phi=0$ the function $U=0$, so this extremum of $V_{eff}$ does not correspond to a de Sitter solution.}.
However,  it is clear that the addition of any monomial $F=F_0\phi^n$ with
$n>2$ and $F_0>0$ gives us a stable de Sitter solution. Straightforward  calculation shows that
\begin{equation*}
\phi_{dS}^{n-2}=\frac{3\xi^2}{nF_0m^2},
\end{equation*}
and, consequently, the de Sitter solution exists if $n \ne 2$. Using the second derivative of the effective potential, we easily obtain
\begin{equation}
    V_{eff}''(\phi) = \frac{4 \xi^2}{m^2} (n-2),
\end{equation}
which implies that  the de Sitter solution is unstable for $n<2$.
Looking at the plot of the effective potential in Fig.~\ref{Figpol2},
one can clearly distinguish between cases corresponding to $n>2$ $\&$ $n<2$.

\subsection{Models with the Higgs potential}
\begin{figure}[h!tbp]
\includegraphics[width=0.72\linewidth]{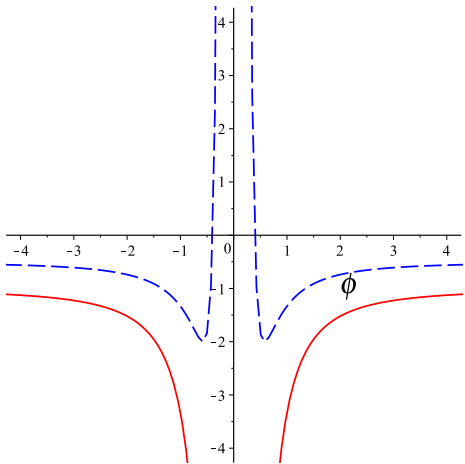} \
\caption{The effective potential $V_{eff}(\phi)$ for $U=1+\phi^2$, $V=V_0\phi^4$, $F=\phi^{-4}$ at
 $V_0=1$ (red solid curve) and at $V_0=2$ (blue dash curve).
\label{Figpol4}}
\end{figure}

Let us consider model with
\begin{equation}
U=U_0+{\xi}{\phi}^{2}\,,\qquad V=V_0{\phi}^{4},
\end{equation}
where $U_0$, $\xi$ and $V_0$ are positive constants.
The corresponding model without the Gauss--Bonnet  term is the physically motivated
inflationary model dubbed  Higgs-driven inflation~\cite{Higgs} perfectly consistent with cosmological observations.
The inflationary scenario proposed in~Ref.~\cite{vandeBruck:2015gjd} uses the functional form of $F$ given by $F= F_0/\phi^4$.
In this case, the effective potential has the following form:
\begin{equation}
V_{eff}={}-\frac { \left( U_0+\xi\phi^{2} \right)^2}{
V_0\phi^{4}}+\frac{2F_0}{3\phi^4}.
\end{equation}
The de Sitter solutions correspond to real values of $\phi$ only if $F_0>3U_0^2/(2V_0)$:
\begin{equation}
\phi_{dS}=\pm{\frac{\sqrt { 2F_0V_0 -3U_0^2}}{\sqrt{3\xi U_0}}}.
\end{equation}
At the de Sitter points, the second derivative of the effective potential is
\begin{equation}
V^{\prime\prime}_{eff}(\phi_{dS})={\frac {72U_0^3{\xi}^{3}}{\left(2F_0V_0-3U_0^2\right)^{2}V_0}}>0.
\end{equation}
Thus, evidently, all de Sitter solutions of the considered model are stable. In Fig.~\ref{Figpol4},
the red solid curve corresponds
to a model without a de Sitter solution, whereas the blue dash curve corresponds to a model with de Sitter solutions.

In another interesting case of the function
\begin{equation}
F=\frac {3U_0^2}{2V_0{\phi}^{4}}+3{\frac{U_0\xi}{
V_0{\phi}^{2}}}+\xi_2{\phi}^{2}+\xi_3{\phi}^{4},
\end{equation}
extrema  of the effective potential are $\phi_{1,2}=\pm \sqrt{-\xi_2/(2\xi_3)}$ and $\phi_0=0$.
The point $\phi_0=0$ is a singular point of the function $F$ and we do not consider this point.
Evidently,  the de Sitter points $\phi_{1,2}$ are real only if the parameters $\xi_3$ and $\xi_2$ have different signs.

In the points $\phi_{1,2}$ the second derivative is $V^{\prime\prime}_{eff}={}-8\xi_2/3$.
So, the points $\phi_{1,2}$ are unstable at $\xi_2>0$ and stable at $\xi_2<0$.

\subsection{Six degree potential}

In this subsection, we consider more general functions $U$ and $V$, than in the previous subsections, and power-law functions $F$:
 \begin{equation}\label{polynomodel}
 U=U_0(1+\xi\phi^N),\quad V=V_0\phi^n, \quad F=F_0\phi^\alpha,
\end{equation}
where $U_0$, $V_0$ and $\xi$ are positive constants\footnote{A similar model has been considered in~\cite{Granda:2017oku}}.
 The effective potential of the model has the following form:
\begin{equation}
V_{eff}={}-{\frac {U_0^{2} \left( 1+\xi\,{\phi}^{N} \right) ^{2}}{V_0{\phi}^{n}}}+\frac23{F_0}{\phi}^{\alpha}.
\end{equation}
In the case $n>2N$, at $F_0=0$ the function $V_{eff}$ is a monotonically increasing one, so there is no de Sitter solution for $U>0$ that has been mentioned in~\cite{Sami:2012uh}.

\begin{figure}[h!tbp]
\includegraphics[width=0.72\linewidth]{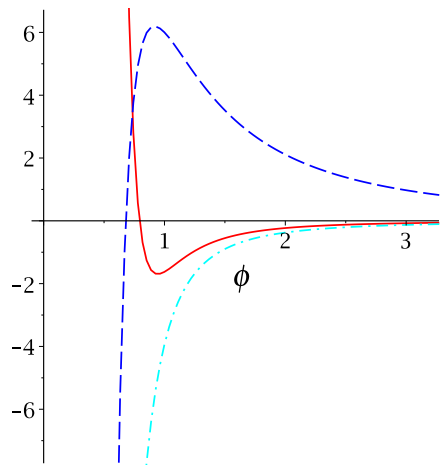}
\caption{The effective potential $V_{eff}(\phi)$ for $U=1+\xi\phi^2$, $V=\phi^6$, $F=F_0\phi^\alpha$.
 Values of parameters are $\alpha=-8$, $\xi=0.72$,  and $F_0=2$  (red solid curve), $\alpha=-2$, $\xi=1$, and $F_0=15$ (blue dash curve), $\alpha=-2$, $\xi=1$,  and  $F_0=0.1$ (cyan dash-dot curve).
\label{Figpol6}}
\end{figure}

Let us consider the case of $N=2$ and $n=6$. If we add the function $F=F_0/\phi^2$, then an unstable de Sitter can be obtained for some values of parameter (see the blue dash and cyan dash-dot curves in Fig.~\ref{Figpol6}). The analysis of $V_{eff}^\prime$ shows that no more than two values of $\phi_{dS}$ are real, either
\begin{equation}\label{DSsol1}
{\phi_{dS}}_{1,2}=\pm{\frac {\sqrt {3\left( 2{U_0}\xi+\sqrt{U_0^{2}{
\xi}^{2}+2{F_0}{V_0}} \right) {U_0}}}{\sqrt {2{F_0}{V_0}-3U_0^{2}{\xi}^{2}}}}\,,
\end{equation}
or
\begin{equation}\label{DSsol3}
{\phi_{dS}}_{3,4}=\pm{\frac {\sqrt {3\left(2{U_0}\xi-\sqrt{U_0^{2}{
\xi}^{2}+2{F_0}{V_0}} \right) {U_0}}}{\sqrt {2{F_0}{V_0}-3U_0^{2}{\xi}^{2}}}}\,.
\end{equation}

In order to  find out the stable de Sitter solutions, we choose $F=F_0/\phi^8$ and obtain $V_{eff}$, which has minima, see the red solid curve in  Fig.~\ref{Figpol6}.

\section{Conclusion}
In the present paper, we  investigate a homogeneous cosmological dynamics of a nonminimally
coupled scalar field  both with the curvature and with the Gauss--Bonnet term. We are mainly interested in looking for the fixed points of scalar field dynamics which correspond to de Sitter solutions. We  show that, in the case of a positive coupling function $U(\phi)$, it is possible to introduce an effective potential $V_{eff}$ which can be expressed
through curvature $U$, the scalar field potential $V$, and the coupling function with the Gauss--Bonnet term denoted by~$F$. We show that it is convenient to investigate the structure of fixed points using
the effective potential, indeed, the stable de Sitter solutions correspond to  minima of the effective potential.
The latter implies that the  existence and stability of de Sitter solutions in the system under consideration can
be studied with the help of function $V_{eff}$, since the stability of de Sitter solutions is analogous to the stability
of a classical mechanical system moving in the potential field $V_{eff}$. One can get the structure and stability properties of de Sitter solutions using  only graphical representation of the effective potential.

 In the model under consideration, the effective potential is a sum of two terms. The first one includes the contributions from $U$ and $V$ --- the effective potential for models in absence of the Gauss--Bonnet term, proposed in~\cite{Skugoreva:2014gka}; the second term includes the function $F$ or the  contribution from the Gauss--Bonnet term. Thus, the use of the effective potential is the simplest way to compare the  results on existence and stability of de Sitter solutions in models with and without the Gauss--Bonnet term.
We in general derive the structure of the effective potential and conditions for existence and stability of de Sitter solutions.
Using this approach, we have studied concrete  models with the Gauss--Bonnet term and described a number
of situations where de Sitter solutions exist due to the presence of the Gauss--Bonnet term and disappear otherwise.

Let us note that the effective potential for models without the Gauss--Bonnet term has a simple physical meaning: it is an invariant under the conformal rescaling of the metric that coincides with the potential in the Einstein frame~\cite{Jarv:2015kga}. It would be interesting to get a similar interpretation for the effective potential proposed in this paper.

Let us also emphasize that the knowledge of unstable de Sitter solutions can be useful to describe inflation (see, for example,~\cite{Elizalde:2014xva} for details), whereas stable de Sitter solutions are often used in models of late time acceleration of the Universe. We should, however, stress that the effective potential formalism in the Gauss--Bonnet cosmology works
only for the study of an exact de Sitter solution. When we deviate from an exact de Sitter solution to a quasi--de Sitter one, the effective
potential as a single construction from three functions, entering the action of the theory, may  not be enough.
We leave the detailed study of quasi--de Sitter solutions in the considered models  for our future investigations.

\section*{Acknowledgement}
This work is partially supported by the Indo--Russia Project; E.P., A.T., and S.V. are supported by RFBR Grant No.~18-52-45016, and M.S. is supported by Grant No.~INT/RUS/RFBR/P-315 of the Department of science and technology of India government. A.T.~is supported by the Russian Government Program of Competitive Growth of Kazan Federal University.

\end{document}